\documentstyle[12pt]{article}

  \def\pp{{\mathchoice
          {%displaystyle
              \kern 1pt%
              \raise 1pt
              \vbox{\hrule width5pt height0.4pt depth0pt
                    \kern -2pt
                    \hbox{\kern 2.3pt
                          \vrule width0.4pt height6pt depth0pt
                          }
                    \kern -2pt
                    \hrule width5pt height0.4pt depth0pt}%
                    \kern 1pt
           }
            {%textstyle
              \kern 1pt%
              \raise 1pt
              \vbox{\hrule width4.3pt height0.4pt depth0pt
                    \kern -1.8pt
                    \hbox{\kern 1.95pt
                          \vrule width0.4pt height5.4pt depth0pt
                          }
                    \kern -1.8pt
                    \hrule width4.3pt height0.4pt depth0pt}%
                    \kern 1pt
            }
            {%scriptstyle
              \kern 0.5pt%
              \raise 1pt
              \vbox{\hrule width4.0pt height0.3pt depth0pt
                    \kern -1.9pt  %[e]=0.15pt
                    \hbox{\kern 1.85pt
                          \vrule width0.3pt height5.7pt depth0pt
                          }
                    \kern -1.9pt
                    \hrule width4.0pt height0.3pt depth0pt}%
                    \kern 0.5pt
            }
            {%scriptscriptstyle
              \kern 0.5pt%
              \raise 1pt
              \vbox{\hrule width3.6pt height0.3pt depth0pt
                    \kern -1.5pt
                    \hbox{\kern 1.65pt
                          \vrule width0.3pt height4.5pt depth0pt
                          }
                    \kern -1.5pt
                    \hrule width3.6pt height0.3pt depth0pt}%
                    \kern 0.5pt%}
            }
        }}

  \def\mm{{\mathchoice
                       {%displaystyle
                             \kern 1pt
               \raise 1pt    \vbox{\hrule width5pt height0.4pt depth0pt
                                  \kern 2pt
                                  \hrule width5pt height0.4pt depth0pt}
                             \kern 1pt}
                       {%textstyle
                            \kern 1pt
               \raise 1pt \vbox{\hrule width4.3pt height0.4pt depth0pt
                                  \kern 1.8pt
                                  \hrule width4.3pt height0.4pt depth0pt}
                             \kern 1pt}
                       {%scriptstyle
                            \kern 0.5pt
               \raise 1pt
                            \vbox{\hrule width4.0pt height0.3pt depth0pt
                                  \kern 1.9pt
                                  \hrule width4.0pt height0.3pt depth0pt}
                            \kern 1pt}
                       {%scriptscriptstyle
                           \kern 0.5pt
             \raise 1pt  \vbox{\hrule width3.6pt height0.3pt depth0pt
                                  \kern 1.5pt
                                  \hrule width3.6pt height0.3pt depth0pt}
                           \kern 0.5pt}
                       }}

\catcode`@=11
\def\un#1{\relax\ifmmode\@@underline#1\else
        $\@@underline{\hbox{#1}}$\relax\fi}
\catcode`@=12

\let\du=\du                     % dot-under
\def\a{\alpha}
\def\b{\beta}

\def\d{\delta}

\def\f{\phi}
\def\g{\gamma}
\def\h{\eta}

\def\k{\kappa}
\def\l{\lambda}
\def\m{\mu}
\def\n{\nu}
\def\o{\omega}
\def\p{\pi}
\def\q{\theta}

\def\x{\xi}

\def\O{\Omega}

\def\S{\Sigma}

\def\bo{{\raise-.5ex\hbox{\large$\Box$}}}               % D'Alembertian
\def\pa{\partial}                                       % curly d
\def\de{\nabla}                                         % del
\def\pr{\prod}                                          % product
\def\TH{{\raise.2ex\hbox{$\displaystyle \bigodot$}\mskip-4.7mu \llap H \;}}
\def\face{{\raise.2ex\hbox{$\displaystyle \bigodot$}\mskip-2.2mu \llap {$\ddot
        \smile$}}}                                      % happy face
\def\dg{\sp\dagger}                                     % hermitian conjugate
\def\sp#1{{}^{#1}}                              % superscript (unaligned)
\def\VEV#1{\left\langle #1\right\rangle}        % < >
\def\leftrightarrowfill{$\mathsurround=0pt \mathord\leftarrow \mkern-6mu
        \cleaders\hbox{$\mkern-2mu \mathord- \mkern-2mu$}\hfill
        \mkern-6mu \mathord\rightarrow$}
\def\dvec#1{\vbox{\ialign{##\crcr
        \leftrightarrowfill\crcr\noalign{\kern-1pt\nointerlineskip}
        $\hfil\displaystyle{#1}\hfil$\crcr}}}           % <--> accent
\def\frac#1#2{{\textstyle{#1\over\vphantom2\smash{\raise.20ex
        \hbox{$\scriptstyle{#2}$}}}}}                   % fraction
\def\sfrac#1#2{{\vphantom1\smash{\lower.5ex\hbox{\small$#1$}}\over
        \vphantom1\smash{\raise.4ex\hbox{\small$#2$}}}} % alternate fraction
\def\bfrac#1#2{{\vphantom1\smash{\lower.5ex\hbox{$#1$}}\over
        \vphantom1\smash{\raise.3ex\hbox{$#2$}}}}       % "
\def\afrac#1#2{{\vphantom1\smash{\lower.5ex\hbox{$#1$}}\over#2}}    % "
\def\[{\lfloor{\hskip 0.35pt}\!\!\!\lceil}
\def\]{\rfloor{\hskip 0.35pt}\!\!\!\rceil}

\def\du#1#2{_{#1}{}^{#2}}
\def\ud#1#2{^{#1}{}_{#2}}

\def\fracm#1#2{\hbox{\large{${\frac{{#1}}{{#2}}}$}}}
\def\ha{{\fracmm12}}
\def\tr{{\rm tr}}

\def\un{\underline}
\def\fracmm#1#2{{{#1}\over{#2}}}

\def\low#1{{\raise -3pt\hbox{${\hskip 0.75pt}\!_{#1}$}}}

\newskip\humongous \humongous=0pt plus 1000pt minus 1000pt

\newif\ifdtup

\def\plpl{\raise-2pt\hbox{$\raise3pt\hbox{$_+$}\hskip-6.67pt\raise0.0pt}}
\def\mimi{\raise-2pt\hbox{$\raise3pt\hbox{$_-$}\hskip-6.67pt\raise0.0pt}}

\def\dvm{\raisebox{-.145ex}{\rlap{$=$}}}
\def\DM{{\scriptsize{\dvm}}~~}

\def\ref#1{$\sp{#1)}$}

\def\pl#1#2#3{Phys.~Lett.~{\bf {#1}B} (19{#2}) #3}
\def\np#1#2#3{Nucl.~Phys.~{\bf B{#1}} (19{#2}) #3}
\def\prl#1#2#3{Phys.~Rev.~Lett.~{\bf #1} (19{#2}) #3}
\def\pr#1#2#3{Phys.~Rev.~{\bf D{#1}} (19{#2}) #3}
\def\cqg#1#2#3{Class.~and Quantum Grav.~{\bf {#1}} (19{#2}) #3}

\def\mpl#1#2#3{Mod.~Phys.~Lett.~{\bf A{#1}} (19{#2}) #3}

\def\ibid#1#2#3{{\it ibid.}~{\bf {#1}} (19{#2}) #3}

\parskip=\medskipamount                 % space between paragraphs (LaTeX)
\thicklines                         % thick straight lines for pictures (LaTeX)

\begin{document}

% =========================== UH title page ==========================

\thispagestyle{empty}               % no heading or foot on title page (LaTeX)

\def\border{                                            % UH border
        \setlength{\unitlength}{1mm}
        \newcount\xco
        \newcount\yco
        \xco=-24
        \yco=12
        \begin{picture}(140,0)
        \put(-20,11){\tiny Institut f\"ur Theoretische Physik Universit\"at
Hannover~~ Institut f\"ur Theoretische Physik Universit\"at Hannover~~
Institut f\"ur Theoretische Physik Hannover}
        \put(-20,-241.5){\tiny Institut f\"ur Theoretische Physik Universit\"at
Hannover~~ Institut f\"ur Theoretische Physik Universit\"at Hannover~~
Institut f\"ur Theoretische Physik Hannover}
        \end{picture}
        \par\vskip-8mm}

\def\headpic{                                           % UH heading
        \indent
        \setlength{\unitlength}{.8mm}
        \thinlines
        \par
        \begin{picture}(29,16)
        \put(75,16){\line(1,0){4}}
        \put(80,16){\line(1,0){4}}
      \put(85,16){\line(1,0){4}}
        \put(92,16){\line(1,0){4}}

        \put(85,0){\line(1,0){4}}
        \put(89,8){\line(1,0){3}}
        \put(92,0){\line(1,0){4}}

        \put(85,0){\line(0,1){16}}
        \put(96,0){\line(0,1){16}}
        \put(92,16){\line(1,0){4}}

        \put(85,0){\line(1,0){4}}
        \put(89,8){\line(1,0){3}}
        \put(92,0){\line(1,0){4}}

        \put(85,0){\line(0,1){16}}
        \put(96,0){\line(0,1){16}}
        \put(79,0){\line(0,1){16}}
        \put(80,0){\line(0,1){16}}
        \put(89,0){\line(0,1){16}}
        \put(92,0){\line(0,1){16}}
        \put(79,16){\oval(8,32)[bl]}
        \put(80,16){\oval(8,32)[br]}

        \end{picture}
        \par\vskip-6.5mm
        \thicklines}

\border\headpic {\hbox to\hsize{
\vbox{\noindent DESY ~96 --182 \hfill  August 1996 \\
ITP--UH--18/96 \hfill  hep-th/9609004 \\ }}}

\noindent
\vskip1.3cm
\begin{center}

{\Large\bf  The $OSp(32|1)$ versus $OSp(8|2)$ supersymmetric 
\vglue.1in 
              M-brane action from self-dual (2,2) strings}~\footnote{Supported 
in part by the `Deutsche Forschungsgemeinschaft' and the `Volkswagen Stiftung'}\\
\vglue.3in

Sergei V. Ketov \footnote{
On leave of absence from:
High Current Electronics Institute of the Russian Academy of Sciences,
\newline ${~~~~~}$ Siberian Branch, Akademichesky~4, Tomsk 634055, Russia}

{\it Institut f\"ur Theoretische Physik, Universit\"at Hannover}\\
{\it Appelstra\ss{}e 2, 30167 Hannover, Germany}\\
{\sl ketov@itp.uni-hannover.de}
\end{center}
\vglue.2in
\begin{center}
{\Large\bf Abstract}
\end{center}

\noindent 
Taking the (2,2) strings as a starting point, we discuss the equivalent integrable
field theories and analyze their symmetry structure in $2+2$ dimensions from the 
viewpoint of string/membrane unification. Requiring the `Lorentz' invariance and 
supersymmetry in the (2,2) string target space leads to an extension of the (2,2)
string theory to a theory of $2+2$ dimensional supermembranes ({\it M-branes}) 
propagating in a higher dimensional target space. The origin of the hidden target
space dimensions of the M-brane is related to the {\it maximally} extended 
supersymmetry implied by the `Lorentz' covariance and dimensional reasons. The 
K\"ahler-Chern-Simons-type action describing the self-dual gravity in $2+2$ 
dimensions is proposed. Its maximal supersymmetric extension (of the 
Green-Schwarz-type) naturally leads to the $2+10$ (or higher) dimensions 
for the M-brane target space. The proposed $OSp(32|1)$ supersymmetric action 
gives the pre-geometrical description of M-branes, which may be useful for a 
fundamental formulation of F\&M theory.  

\newpage

{\bf 1} {\it Introduction.} Since a discovery of string dualities, much evidence 
was collected for the idea that `different' string theories can be understood as 
particular limits of a unique underlying theory whose basic formulation is yet
to be found. The fundamental theory does not seem to be a theory of strings but 
it describes fields, strings and membranes in a democratic way. A candidate for 
the unified theory was also proposed under the name of {\it M-theory}~\cite{sch1,
w1} or its refined {\it F-theory} formulation~\cite{va1}, which can be reduced 
to all known 10-dimensional supertrings and 11-dimensional supergravity as well. 
There should be also room for strings with {\it extended}  world-sheet 
supersymmetry in the unified theory. The anticipated relation between the 
$N=(2,1)$ heterotic strings and M-theory in some particular low-dimensional 
backgrounds was recently used~\cite{kmo} to propose the definition of the 
underlying M-theory as a theory of $2+2$ dimensional membranes (called 
{\it M-branes}~\cite{town1}) embedded in higher dimensions. The origin of 
M-branes should therefore be understood from the basic properties of N=2 strings.
It is the purpose of this Letter to argue that the hidden membrane (both 
world-volume and target space) dimensions are in fact {\it required} by natural 
symmetries which are broken in the known N=2 string formulations. By the 
{\it natural} symmetries I mean `Lorentz' invariance and supersymmetry which 
should be made explicit and linearly realized. That symmetries uniquely determine 
the dynamics of M-branes.

The basic idea for describing M-branes naturally arises from the known
{\it world-sheet/target space duality} of N=2 strings. In the early days of N=2 
string theory, when only two-dimensional target spaces were considered, 
Green~\cite{green} suggested to use the N=2 string world-sheet as the target 
space, which implies a duality between the world-sheet moduli and their target 
space counterparts. The four-dimensional nature of the $(2,2)$ string target 
space as a (hyper) K\"ahler manifold was understood later by Ooguri and 
Vafa~\cite{ov}, who suggested to associate with the N=2 string world-sheet 
(Riemann surface) a four-dimensional symplectic space -- the so-called `cotangent
bundle of the Riemann surface'. The latter has an equal number of moduli to be
associated with non-trivial deformations of a complex structure {\it and} of a 
K\"ahler class. The duality (in fact, triality) symmetries then appear between 
the world-sheet moduli, the target space complex structure moduli, and the target
space K\"ahler-class moduli. I am going to use the world-sheet/target space 
duality of N=2 strings as the (first) working  principle of string/membrane 
unification, namely, as a route for constructing the self-dual theory of M-branes
 out of the target space field theory of (2,2) strings, along the lines of
ref.~\cite{kmo}. However, unlike the way of reasoning in ref.~\cite{kmo}, which 
puts forward the (2,1) heterotic strings, I consider closed and open (2,2) 
strings as a starting point. The critical (2,2) strings naturally live in $2+2$ 
dimensions, which are crucial for self-duality, whereas the $(2,0)$ or $(2,1)$ 
heterotic strings require the $1+1$ or $1+2$ dimensional target space~\cite{ov}. 
The duality principle is however not enough to deliver the M-brane action, since 
it does not say enough about the symmetries of the M-brane. Hence, I postulate
the second working principle by requiring all the natural symmetries to explicitly
 appear in the target space action. Despite its innocent content, the `Lorentz' 
invariance in $2+2$ dimensions appears to be non-trivial for N=2 strings. By 
gauging the `Lorentz' group $SO(2,2)$, I formulate a K\"ahler-Chern-Simons-type
gauge-invariant action in five dimensions, whose dynamics describes the self-dual
gravity in four dimensions. It gives the relevant part of the M-brane action, 
according to the (first) duality principle above. The rest of the M-brane action 
is fixed by requiring the maximal supersymmetry (the second working principle) in
the M-brane target space, whose dimension is $2+10$, or it can be even higher. 
The supersymmetric action is proposed to describe the M-branes, which may be
the fundamental constituents of the putative F\&M theory.

{\bf 2} {\it Summary of (2,2) strings}. The N=2 strings are strings with {\it two}
 world-sheet (local) supersymmetries.~\footnote{See refs.~\cite{markus,book} for 
a review.} The gauge-invariant $N=2$ string world-sheet actions in the NSR-type 
formulation are given by couplings of a two-dimensional N=2 supergravity to a 
complex N=2 scalar matter~\cite{bs}, and they possess global 
$U(1,1)\times {\bf Z}_2$ target space 
symmetry. A covariant gauge-fixing introduces conformal ghosts $(b,c)$, complex 
superconformal ghosts $(\b^{\pm},\g^{\mp})=(\pa\x^{\pm}e^{-\f^{\mp}},\h^{\mp}
e^{\f^{\mp}})$, and real abelian ghosts $(\tilde{b},\tilde{c})$, as usual. The 
chiral N=2 (superconformal) current algebra comprises a stress-tensor $T(z)$, two
supercurrents $G^{\pm}(z)$, and an abelian current $J(z)$. The critical closed 
and open (2,2) strings live in four dimensions with a signature $2+2$.~\footnote{
The signature is dictated by the (2,2) world-sheet supersymmetry. The euclidean 
signature is \newline ${~~~~~}$ excluded by trivial kinematics for massless 
particles.} The current algebras of the N=2 heterotic strings have the additional
abelian {\it null} current. It is needed for a nilpotency of the BRST charge, and
implies a reduction of the N=2 string target spacetime dynamics down to $1+2$ or 
$1+1$ dimensions~\cite{ov}.
 
The BRST cohomology and on-shell amplitudes of N=2 strings were investigated by 
several groups~\cite{ov,bov,te,ha}. There exists only a single massless 
physical state in the open or closed (2,2) string spectrum. This particle can be 
identified with the {\it Yang} scalar of self-dual Yang-Mills theory for open 
strings, or the {\it K\"ahler} scalar of self-dual supergravity for closed 
strings, while infinitely many massive string modes are all unphysical. The (2,2)
strings thus lack `space-time' supersymmetry. Though twisting the N=2 
superconformal algebra yields some additional twisted physical states  which 
would-be the target space `fermions', they actually decouple. It is consistent 
with another observation that the `space-time fermionic' vertex operators 
constructed in ref.~\cite{ha} anticommute modulo picture-changing, instead of 
producing `space-time' translations required by the `space-time' supersymmetry. 

An $n$-point function of closed N=2 strings is given by a topological sum 
indexed by the genus $g$ and the instanton number (Chern class) $c$, with each 
term in the sum being an integral over metric, N=2 fermionic and Maxwell moduli. 
The Maxwell
moduli parameterize the space of flat connections or harmonic 1-forms $H$ on the 
$n$-punctured world-sheet $\S$, and they enter the gauge-fixed action as 
$\int_{\S}H\wedge *J$. Making a shift $H\to H+h$ changes the  action as
$$\int_{\S}h\wedge *J=\sum^{g}_{i=1}\left(\int_{a_i}h\oint_{b_i}*J  -
\int_{b_i}h\oint_{a_i}*J\right)+\sum^n_{l=1}\oint_{c_l}h\oint_{p_0}^{p_l}*J~,$$
where a canonical homology basis $(a_i,b_j)$ on $\S$, the contours $c_l$ 
encircling punctures $p_l$, and a reference point $p_0$ have been introduced. 
Therefore, the shift gives rise to twists 
$SFO(\q)\equiv\exp\left\{2\p i\q\int *J\right\}$ 
around the homology cycles as well as around the punctures, with $\q\in \[0,1\]$.
This phenomenon is known as {\it spectral flow}. A twist around a puncture at $z$
can be absorbed into a redefined (twisted) vertex operator~\cite{ha}
$$ V(z)\to V^{(\q)}(z)=\exp\left\{2\p i\q\int^{z}_{z_0} *J\right\}V(z)~.\eqno(1)$$
The spectral flow operator $SFO$ is BRST-closed but only its zero mode is not 
BRST-exact. Hence, the position of $SFO$ in an amplitude is irrelevant, and all 
the $n$-point functions are invariant,
$$ \VEV{V_1^{(\q_1)}\cdots V_n^{(\q_n)}}=\VEV{V_1\cdots V_n}~,\eqno(2)$$
as long as the total twist vanishes, $\sum_l \q_l=0$. The bosonized spectral flow
operator reads
$$ SFO(\q)=e^{-2\p i\q\f(z_0)}\exp\left\{ 2\p i\q\f(z)\right\}~,\eqno(3)$$
where the $U(1)$ current has been bosonized as $*J=d\f$. The two factors in 
eq.~(3) are separately neutral under the local $U(1)$, but carry opposite charges
under the {\it global} $U(1)$ symmetry. Eq.~(3) relates the spectral flow to 
Maxwell instantons on the world-sheet. Indeed, choosing $\q=1$ yields an 
{\it instanton-creation operator}, $ICO\equiv \l\, SFO(\q=1)$, which changes the 
world-sheet instanton number $c$ by one. Amplitudes with different instanton 
backgrounds are therefore related as
$$ \VEV{V_1\cdots V_n}_c=\VEV{V_1\cdots V_n(ICO)^c}_{c=0}
=\l^c\VEV{V_1^{(\q_1)}\cdots V_n^{(\q_n)}}_{c=0}~,\eqno(4)$$
with a total twist of $\sum_l \q_l=c$. Hiding the reference point ambiguity by
declaring the Maxwell coupling constant to be $\l=\exp\left\{ 2\p i\f(z_0)
\right\}$ implies that both $ICO$ and $\l$ have a {\it non-vanishing} charge with
respect to the $U(1)$ subgroup of the actual global symmetry group 
$U(1,1)\subset SO(2,2)$~\cite{ha}.

The only non-vanishing N=2 string scattering amplitudes are 3-point trees (and, 
maybe, 3-point loops as well), while all the other tree and loop  amplitudes 
vanish due to kinematical reasons.~\footnote{Similar results are valid for open 
(2,2) strings too.} As a result, a (2,2) string theory appears to be equivalent to
an {\it integrable} field theory. In particular, the open (2,2) string amplitudes 
are reproduced by either the {\it Yang} non-linear sigma-model action~\cite{yang}
or the {\it Leznov-Parkes} cubic action~\cite{lp}, each following from a field 
integration of the self-dual Yang-Mills (SDYM) equations in a particular gauge, 
and related to each other by a duality transformation. As far as the closed (2,2) 
strings in the zero-instanton sector are concerned, the equivalent non-covariant 
field theory action is known as the {\it Pleba\'nski}  action~\cite{ple} for the 
self-dual gravity (SDG). The world-sheet instanton effects lead to {\it a 
deformation} of self-duality: the Ricci-tensor does not vanish, while the 
integrability implies the self-dual {\it Weyl\/} tensor instead. 

The natural (global) `Lorentz' symmetry of a flat $2+2$ dimensional `space-time'
is $SO(2,2)\cong SU(1,1) \otimes SU(1,1)'$. The NSR-type N=2 string actions used 
to calculate the amplitudes have only a part of it, namely, $SU(1,1)$, so is the 
symmetry of the N=2 string amplitudes. The full `Lorentz' symmetry $SO(2,2)$ can
be formally restored in the {\it twistor} space, which adds the (harmonic) space 
$SU(1,1)'/U(1)$ of all complex structures in $2+2$ dimensions~\cite{bov}. 
The ladder generators of the second $SU(1,1)'$ factor can be explicitly 
constructed as follows~\cite{bov}:
$$J_-=\int \x^-\h^-(1-c\tilde{b})ICO~,\qquad 
J_+=\int \x^+\h^+(1+c\tilde{b})ICO^{-1}~.\eqno(5) $$
Closing the underlying N=2 superconformal algebra to be appended by the 
additional currents $J_{\pm}$ results in the so-called `small'\/ {\it twisted\/} 
N=4 superconformal algebra. This remarkable property allows one to treat the N=2 
string theory as an N=4 {\it topological} field theory~\cite{bov,gs}. The 
embeddings of the N=2 algebra into the N=4 algebra are just parameterized by 
twistors: a choice of a complex structure selects a $U(1,1)$ subgroup of the
`Lorentz' group, while world-sheet Maxwell instantons rotate that complex 
structure. 

The (real) coupling constant $g$ of the (2,2) string interaction and the Maxwell 
coupling constant (phase) $\l$ can be naturally unified into a single complex 
coordinate parameterizing the moduli space of complex structures. The complex N=2
string coupling can also be interpreted as the vacuum expectation value of a {\it
complex dilaton} field. Hence, the N=2 string dilaton is {\it not} inert under 
the full `Lorentz' transformations~! The dilaton thus takes its values in 
$SU(1,1)'/U(1)'$, and it can therefore be represented by an {\it anti-self-dual} 
(closed) two-form $\o$ satisfying a {\it nilpotency} condition $\o\wedge\o=0$ 
(see sects.~4 and 5 also). 

{\bf 3} {\it Adding supersymmetry in $2+2$ dimensions}. Because of the 
isomorphisms $SU(1,1)\cong SL(2,{\bf R})$ and $SO(2,2)\cong SL(2,{\bf R})\otimes 
SL(2,{\bf R})'$, it is natural to represent the $2+2$ `space-time' coordinates
as $x^{\a,\a'}$, where $\a=(+,-)$ and $\a'=(+',-')$ refer to $SL(2)$ and $SL(2)'$,
respectively. The $N$-extended supersymmetrization of self-duality amounts to 
extending the $SL(2)$ factor to $OSp(N|2)$, while keeping the $SL(2)'$ one to 
be intact. One has $\d^{AB}=(\d^{ab},C^{\a\b})$, where $\d^{ab}$ is the $SO(N)$ 
metric and $C^{\a\b}$ is the (part of) charge conjugation matrix, $A=(a,\a)$. 
In superspace $Z=(x^{\a,\a'},\q^{A'})$, the $N$-extended (gauged) self-dual 
supergravity (SDSG) is defined by the constraints on the spinorial covariant 
derivatives, $\de_{A\a'}=E\du{A\a'}{M\m'}\pa_{M\m'} 
+\ha \O_{A\a'BC}M^{CB}$, as~~\cite{sie}:
$$ \{ \de^{a\a},\de^{b\b}\}=C^{\a\b}M^{ab}+\d^{ab}M^{\a\b}~,\eqno(6a)$$
$$\{ \de^{a\a},\de_{b\b'}\}=\d^{a}_{b}C^{\a\b}\de_{\b\b'}~,\quad  
\[ \de^{a\a},\de_{\b\b'}\]=\d^{\a}_{\b}\d^{ab}\de_{b\b'}~,\eqno(6b)$$
where $M^{AB}=(M^{ab},M^{\a\b},\de^{a\a})$ are the generators of $OSp(N|2)$. 
Eqs.~(6) have the $OSp(N|2)\otimes SL(2)'$ (local$\otimes$global) symmetry, 
and they can be `solved' in the light-cone gauge in terms of a SDSG 
{\it pre-potential}. It is well-known that, as far as the SDG is concerned, one 
has
$$ R\low{\a\low{1}\a\low{2}\a\low{3}\a\low{4}}\sim\pa\low{\a\low{1}+'}\pa\low{
\a\low{2}+'}\,g\low{\a\low{3}\a\low{4}-'-'}\sim \pa\low{\a\low{1}+'}\pa\low{
\a\low{2}+'}\pa\low{\a\low{3}+'}\pa\low{\a\low{4}+'}V\low{\DM'\DM'}~~, \eqno(7)$$
where the prepotential $V_{\DM'\DM'}$ has a single component representing the
helicity (+2). Eq.~(7) can be generalized in superspace, 
$R\low{A_1A_2A_3A_4}(Z)\sim\pa\low{A_1+'}\cdots\pa\low{A_4+'}V\low{\DM'\DM'}(Z)$,
where $V\low{\DM'\DM'}$ is a SDSG pre-potential of dimension $(-1)$. The free 
field equation for the SDSG pre-potential, 
$\pa\low{A}{}^{\a'}\pa\low{B\a'}V\low{\DM'\DM'}(Z)=0$, can be solved for all 
$\q^{a-'}$ dependence. It reduces $V\low{\DM'\DM'}(Z)$ to a {\it self-dual} 
superfield $V_{\DM'\DM'}(x^{\a,\a'},\q^{a+'})$, which merely depends on 
{\it a half} of $\q$'s. Of course, it breaks the `Lorentz' symmetry. As a result,
the SDSG constraints in the light cone-gauge can be reduced to a single equation 
for the pre-potential, which is obtained from the $N$-{\it extended\/} 
super-Pleba\'nski action~\cite{sie},
$$S_{\rm SDSG}=\int d^{2+2}x d^N\q\left[ \frac{1}{2} V_{\DM'\DM'}\bo V_{\DM'\DM'}
+\frac{i}{6}V_{\DM'\DM'}(\pa\ud{\a}{+'}\pa_{A+'}V_{\DM'\DM'})
\d^{BA}(\pa_{B+'}\pa_{\a+'}V_{\DM'\DM'})\right]~.\eqno(8)$$
As was noticed by Siegel~\cite{sie}, the action (8) implies the {\it maximal}
supersymmetry~! Indeed, dimensional analysis immediately yields $N=8$, and the 
same follows from counting the total $GL(1)'$ charge of the action (8), where 
$GL(1)'$ is the unbroken part of the `Lorentz' factor $SL(2)'$. Similarly, the 
$N$-{\it extended} super-Leznov-Parkes action for the self-dual supersymmetric 
Yang-Mills (SDSYM) theory implies $N=4$~\cite{sie}:
$$ S_{\rm SDSYM}=\int d^{2+2}xd^4\q \left[ \frac{1}{2} V_{\DM'}\bo V_{\DM'}
+\fracm{i}{3}V_{\DM'}(\pa\ud{\a}{+'}V_{\DM'})(\pa_{\a+'}V_{\DM'})\right]~.
\eqno(9)$$
The SDSG and SDSYM theories in eqs.~(8) and (9) are similar to the non-self-dual 
supersymmetric gauge theories in the light-cone gauge~\cite{lc3}.

The natural appearance of the maximal $N=8$ supersymmetry and the (gauged) 
$SO(8)$ internal symmetry in the supersymmetrized (2,2) string effective action
in $2+2$ dimensions is very remarkable, since that effective action is supposed 
to be a (dual) part of an M-brane action. We may now proceed in the usual way 
known in supergravity, and `explain' the maximally extended local supersymmetry 
as a {\it simple} local supersymmetry in higher dimensions. For example, one may 
use the embedding 
$$ SO(2,2) \otimes SO(8) \subset SO(2,10)~,\eqno(10)$$
which implies going up to $2+10$ dimensions. Indeed, the $2+10$ dimensions are 
the nearest ones in which Majorana-Weyl spinors and self-dual tensors also appear,
 like in $2+2$ dimensions. It should be noticed that twelve dimensions for
string theory were originally motivated in a very different way, namely, by a 
desire to explain the S-duality of type IIB string in ten dimensions as the 
T-duality of a 12-dimensional F-theory dimensionally reduced on a two-torus. The 
type IIB string is then supposed to arise upon double dimensional reduction from 
the F-theory. 

There is, however, a problem with that naive approach. One has to {\it double} 
the on-shell number $(8)$ of the anticommuting coordinates in a covariant M-brane 
action while maintaining the number of their degrees of freedom. One then gets 
$2\times 16=32$ off-shell components, which is just needed for a single 
Majorana-Weyl spinor in $2+10$ dimensions. As is well known in superstring theory,
 it is the $\k$-symmetry of the Green-Schwarz superstring action that makes the 
doubling to be possible, while the Green-Schwarz action itself can be understood
as the particular {\it Wess-Zumino-Novikov-Witten} (WZNW) model with superspace 
as the target supermanifold~\cite{hmu}. Therefore, one should look for a 
{\it Green-Schwarz-type} reformulation of self-duality in $2+2$ dimensions, and 
then maximally supersymmetrize the target space, instead of (or, maybe, 
in addition to) the world-volume (or NSR-type) supersymmetrization. 

{\bf 4}. {\it K\"ahler-Chern-Simons actions for SDYM and SDG}. A more geometrical
(dual) description of the SDYM theory is provided by the {\it five}-dimensional
hyper K\"ahler-Chern-Simons action~\cite{dns}:
$$ S_{\rm hKCS} = -\fracmm{1}{4\p}\int_{Y} \tr\left( \tilde{A}\wedge\tilde{d}
\tilde{A}+\fracm{2}{3}\tilde{A}\wedge\tilde{A}\wedge\tilde{A}\right)\wedge 
\o^ie_i~,\eqno(11)$$
where $Y=M_4\otimes R$, with $M_4$ being the $2+2$ dimensional hyper K\"ahler 
world-volume and $R$ being the auxiliary dimension called extra `time' $t$. Here
$\tilde{A}$ is the Lie algebra valued 1-form on $Y$, $\o^i$ is the hyper K\"ahler
structure on $M_4$, and $e_{\m}=(1,e_i)$ is a basis of quaternions.~\footnote{
There exist the K\"ahler (1,1) form $\o$ and a closed 
(2,0) form $\o^+$ on any hyper K\"ahler manifold \newline ${~~~~~}$ $M_4$. The 
hyper K\"ahler structure is defined by $\o^1={\rm Re}\,\o^+$, 
$\o^2={\rm Im}\,\o^+$ and $\o^3=\o$.} Since $\o^i$ are closed, the action (11) is
invariant under the gauge transformations $\tilde{A}^h=h\tilde{A}h^{-1}-dhh^{-1}$
which should be trivial on the boundary $\pa Y$. I assume that the boundary 
conditions for the gauge field $\tilde{A}$ are chosen in such a way that no 
boundary terms appear in the equations of motion. It is convenient 
to decompose both the gauge field and the exterior derivative into the `time' and
`rest' components, $\tilde{A} =A_t + A$ and $\tilde{d} =dt\fracmm{\pa}{\pa t} +d$.
One finds that $A_t$ and $\o^i$ appear in eq.~(11) as Lagrange multipliers, which
implement the self-duality equations 
$$F\wedge \o^i=0~,\qquad i=1,2,3~,\eqno(12)$$ 
where the YM field strength $F=dA +A\wedge A$ has been introduced. Varying 
with respect to $A$ implies (in the gauge $A_t=0$) that the A-field is 
$t$-independent. In the gauge $A_t=0$, the gauge symmetry is represented by the 
$t$-independent gauge transformations. Therefore, the action (11) describes 
on-shell the SDYM in $2+2$ dimensions. Eq.~(12) for $i=1,2$ can be solved in 
complex coordinates $(z^a,\bar{z}^{\bar{a}})$ on $M_4$ as 
$A_a=(U)^{-1}\pa_a U$ and $A_{\bar{a}}=-\pa_{\bar{a}}U^{\dg}(U^{\dg})^{-1}$, where
$U$ is locally defined. In terms of the gauge-invariant potential $J=UU^{\dg}$, 
the remaining eq.~(12) at $i=3$ is just the {\it Yang} equation,~\footnote{If one
first solves eq.~(12) for $i=2,3$, the remaining equation for $i=1$ follows from 
the dual \newline ${~~~~~}$ Leznov-Parkes action (sect.~2).}
$$ \o\wedge \bar{\pa}\left(J^{-1}\pa J\right)=0~.\eqno(13)$$ 
Eq.~(13) can be obtained from the {\it Donaldson-Nair-Schiff} (DNS) 
action~\cite{dns}
$$ S_{\rm DNS}[J;\o] = -\fracmm{1}{4\p}\int_{M_4}\,\o\wedge\tr(J^{-1}\pa J
\wedge J^{-1}\bar{\pa}J) + \fracmm{i}{12\p}\int_{M_4\times \[0,1\]}\,
\o\wedge\tr(J^{-1}dJ)^3~.\eqno(14)$$

The action similar to eq.~(11) can also be constructed for SDG. Let us simply 
replace the YM Chern-Simons form by the `{\it Lorentz}' Chern-Simons form 
$C_{\rm 3L}$,
$$C_{\rm 3L} = \tr\left( \tilde{\O}\wedge\tilde{d}\tilde{\O}
+\fracm{2}{3}\tilde{\O}\wedge\tilde{\O}\wedge\tilde{\O}\right)=
\tr\left( \tilde{\O}\wedge\tilde{R}-\fracm{1}{3}\tilde{\O}\wedge
\tilde{\O}\wedge\tilde{\O}\right)~,\eqno(15)$$
where $\tilde{R}=\tilde{d}\tilde{\O}+\tilde{\O}\wedge\tilde{\O}$, and the 1-form
$\tilde{\O}$ takes values in the Lie algebra of $SO(2,2)$. The SDG action for a
hyper K\"ahler manifold $M_4$ equipped with the anti-self-dual two-form $\o$ is 
given by
$$ S_{\rm SDG}[\tilde{\O};\o] = -\fracmm{1}{4\p}\int_{Y} C_{\rm 3L}\wedge \o~.
\eqno(16)$$
The self-duality condition $R\wedge\o=0$ appears from varying eq.~(16) with 
respect to $\O_t$ (in the gauge $\O_t=0$). One can check that the vanishing 
variation of the action (16) with respect to a K\"ahler potential to be associated
with $\o$ is consistent with the self-dual geometry, and the physics associated 
with eq.~(16) is four-dimensional indeed. The anti-self-dual two-form $\o$ is 
interpreted as the $(2,2)$ string dilaton field (sect.~2), rather than the 
world-volume gravity.

{\bf 5}. {\it Higher dimensions versus extended supersymmetry}. The maximally
supersymmetric SDSG and SDSYM actions (8) and (9) have manifest $OSp(8|2)$ or 
$OSp(4|2)$ supersymmetry, respectively. In fact, they possess an even larger
{\it superconformal} symmetry $SL(8|4)$ or $SL(4|4)$, respectively~\cite{sie}, 
which may be the fundamental world-volume symmetries of (closed or open) M-branes.
Indeed, the conformal extension of $SO(2,2)$ is given by $SO(3,3)\cong SL(4)$, 
whereas its $N$-supersymmetric extension is just $SL(N|4)$.

Since the internal symmetry of the supergroup $SL(4|4)$ is also  
$SL(4)\cong SO(3,3)$, combining it with the `space-time' conformal group $SO(3,3)$
 implies `hidden' twelve dimensions in yet another way: $SO(3,3)\otimes SO(3,3) 
\subset SO(6,6)$. The $6+6$ dimensions is the only alternative to $2+10$ 
dimensions where Majorana-Weyl spinors also exist. I do not consider this 
possibility.

The gauge actions (11) and (16) for SDYM and SDG, or the equivalent DNS action 
(14), can be naturally supersymmetrized {\it \'a la} Green-Schwarz. The simple 
supersymmetry with {\it one} spinor generator (minimal grading) in the maximal 
dimensions (twelve) amounts to the simple superalgebra $OSp(32|1)$. The choice of
$OSp(32|1)$ is unique since it simultaneoulsy represents the minimal
supersymmetric extension of (i) the (self-dual) `Lorentz' algebra in $2+10$ 
dimensions, (ii) de Sitter algebra in $1+10$ dimensions and (iii) the conformal 
algebra in $1+9$ dimensions~\cite{hp}. A supersymmetry part of $OSp(32|1)$ reads
 ({\it cf.} eq.~(6a)):  
$$\{Q_{\a},Q_{\b}\}=\g^{\m\n}_{\a\b}M_{\m\n} + 
\g_{\a\b}^{\m_1\cdots\m_6}Z_{\m_1\cdots\m_6}^+~,\eqno(17)$$
where $Q_{\a}$ is a 32-component Majorana-Weyl spinor, the Dirac $\g$-matrices are
chirally projected, $M_{\m\n}$ are 66 `Lorentz' generators,~\footnote{At this 
point my approach differs from that of Bars~\cite{bars}.} and 462 
generators $Z_{\m_1\cdots\m_6}^+$ comprise a self-dual six-form (all in $2+10$ 
dimensions). The pre-geometrical action I propose for M-branes is given
by
$$ S_{\rm M}[\tilde{\O};\o] = -\fracmm{1}{4\p}\int_{Y} {\rm str}\left( 
\tilde{\O}\wedge\tilde{d}\tilde{\O}+\fracm{2}{3}\tilde{\O}\wedge\tilde{\O}
\wedge\tilde{\O}\right)\wedge \o~,\eqno(18)$$
where $\o$ is an anti-self-dual two-form (N=2 string dilaton~!) in the 
world-volume, and $\tilde{\O}$ is the $OSp(32|1)$ Lie superalgebra valued 1-form 
gauge potential. The action (18) can be further ({\it doubly}) supersymmetrized 
with respect to the world-wolume, as in sect.~3. Currently, it is unclear to me
 whether it should be done or not.

The action in eq.~(18) is called {\it pre-geometrical} because of the apparent 
absence of the translation generators (momenta) in the gauged superalgebra 
$OSp(32|1)$.~\footnote{A super-Poincar\'e algebra does {\it not} exist in $2+10$ 
dimensions.} However, the momenta can be easily recovered after a 
Wigner-In\"on\"u-type contraction of $OSp(32|1)$ to lower dimensions. 
For instance, 
the 66 Lorentz generators $M_{\m\n}$ are decomposed into 55 Lorentz generators 
and 11 translations in {\it eleven} dimensions. The  additional generators 
$Z_{\m_1\cdots\m_6}^+$ can be interpreted either as the off-shell charges that do
not transform the physical states~\cite{hp}, or as the active charges which are
related to boundaries of extended objects (6-branes)~\cite{bars}. The most
degenerate contraction of $OSp(32|1)$ yields the flat target space in $66+462=528$
(!) dimensions ({\it cf.} ref.~\cite{bars}).

{\bf 6.} {\it Conclusion}. My arguments in this Letter support the 
idea~\cite{kmo} that the fundamental framework for describing the secret
F(or M, S, $\ldots$) theory is provided by the $2+2$ dimensional supermembranes 
(M-branes) living in $2+10$ dimensions. The integrability (or self-duality) of 
M-branes naturally substitutes and generalizes the conformal symmetry of the 
string world-sheet. The basic assumptions were merely the N=2 string 
(world-sheet/target space) duality, and the manifest 'Lorentz' invariance and 
supersymmetry in `space-time'. The hidden superconformal symmetries of M-branes 
are to be responsible for their full 
integrability and the absence of loop divergences in $2+2$ world-volume 
dimensions despite of the fact that the DNS action is non-linear and, hence, is
formally non-renormalizable in four dimensions. It fact, the DNS action is known 
to be {\it one-loop finite}, at least~\cite{yale,ke2}. Its maximally 
supersymmetric extension may have no divergences at all, presumably because of 
having a chiral {\it current} symmetry algebra similar to that in the 
two-dimensional supersymmetric WZNW models~\cite{yale,ke2}. Unlike the N=2 
strings having severe infra-red divergences in loops~\cite{ir}, no such problems 
are expected for M-branes due to the higher world-volume dimension. The theory of
M-branes should therefore exist as a quantum theory, in which strings would 
appear as asymptotic states of M-branes.

\section*{Acknowledgements}

\noindent I am grateful to Olaf Lechtenfeld and Alexander Popov for
useful discussions.

\end{document}